\documentclass[5p,10pt]{elsarticle}

\journal{arXiv}

\usepackage{lineno}

\usepackage{amsmath}
\usepackage{amsfonts}
\usepackage[algo2e,ruled,lined]{algorithm2e} 
\usepackage{booktabs}
\usepackage{multirow}
\usepackage{adjustbox}
\usepackage{ulem}
\usepackage{hyperref}
\usepackage{color}

\usepackage{graphicx}
\graphicspath{{./}{./figs/}}

\newcommand{\edit}[1]{#1}
\newcommand{\sstitle}[1]{\smallskip\noindent\textbf{#1.\/}}

\newtheorem{mydef}{Definition}

\usepackage{xspace}
\newcommand{\toolname}{\texttt{FWHDNN}\xspace}

\begin{document}

\begin{frontmatter}

\title{Handling Heterophily in Recommender Systems with Wavelet Hypergraph Diffusion}

\author{Darnbi Sakong}
\author{Thanh Tam Nguyen}

%
\affiliation{
organization={Griffith University},
country={Australia}
}
%

\begin{abstract}
Recommender systems are pivotal in delivering personalised user experiences across various domains. However, capturing the heterophily patterns and the multi-dimensional nature of user-item interactions poses significant challenges. To address this, we introduce \toolname (Fusion-based Wavelet Hypergraph Diffusion Neural Networks), an innovative framework aimed at advancing representation learning in hypergraph-based recommendation tasks. The model incorporates three key components: (1) a cross-difference relation encoder leveraging heterophily-aware hypergraph diffusion to adapt message-passing for diverse class labels, (2) a multi-level cluster-wise encoder employing wavelet transform-based hypergraph neural network layers to capture multi-scale topological relationships, and (3) an integrated multi-modal fusion mechanism that combines structural and textual information through intermediate and late-fusion strategies. Extensive experiments on real-world datasets demonstrate that \toolname surpasses state-of-the-art methods in accuracy, robustness, and scalability in capturing high-order interconnections between users and items.
\end{abstract}


\end{frontmatter}

\section{Introduction}

Recommender systems often rely on Collaborative Filtering (CF) to predict items that users might prefer by analyzing their past interactions and finding similarities with others who share comparable behaviors. This approach has become essential in improving user experiences across diverse applications, such as social networking platforms~\cite{huang2021knowledge}, streaming services~\cite{liu2021concept}, and online shopping sites~\cite{huang2019online}, by filtering through large volumes of content.

The advent of graph neural networks (GNNs) has inspired recent advancements in representation learning by leveraging the structural properties of user-item interaction data to capture collaborative signals inherent in graph topologies. \edit{GNN-based CF models offer significant advantages for recommendation systems by effectively capturing and representing the intricate relationships between users and items through nodes and edges. This structural representation allows the models to uncover hidden patterns, similarities, and community structures, enhancing the accuracy and relevance of recommendations.}
\edit{For example,} models like LightGCN~\cite{he2020lightgcn} and UltraGCN~\cite{mao2021ultragcn} simplify traditional graph convolutional networks (GCNs) to better align with the needs of recommendation systems.
\edit{Such GNN-based CF models are limited in their ability to capture complex, multi-way relationships, as they primarily focus on pairwise interactions. Furthermore, these models often struggle to represent and leverage higher-order connections, which can result in less nuanced and comprehensive recommendations.}

To further enhance embedding expressiveness with intricate user-item relationships, hypergraphs have been integrated into the CF framework due to their ability to represent complex group-level interactions between users and items. Unlike traditional binary interaction graphs, hypergraphs connect multiple users to a specific item via hyperedges, encapsulating their multi-way interactions. A notable example is SHT~\cite{sht2022}, which employs a integration of  hypergraph transformers to capture global collaborative relationships for recommendation task.

\begin{figure}[!h]
	\centering
	\includegraphics[width=0.6\linewidth]{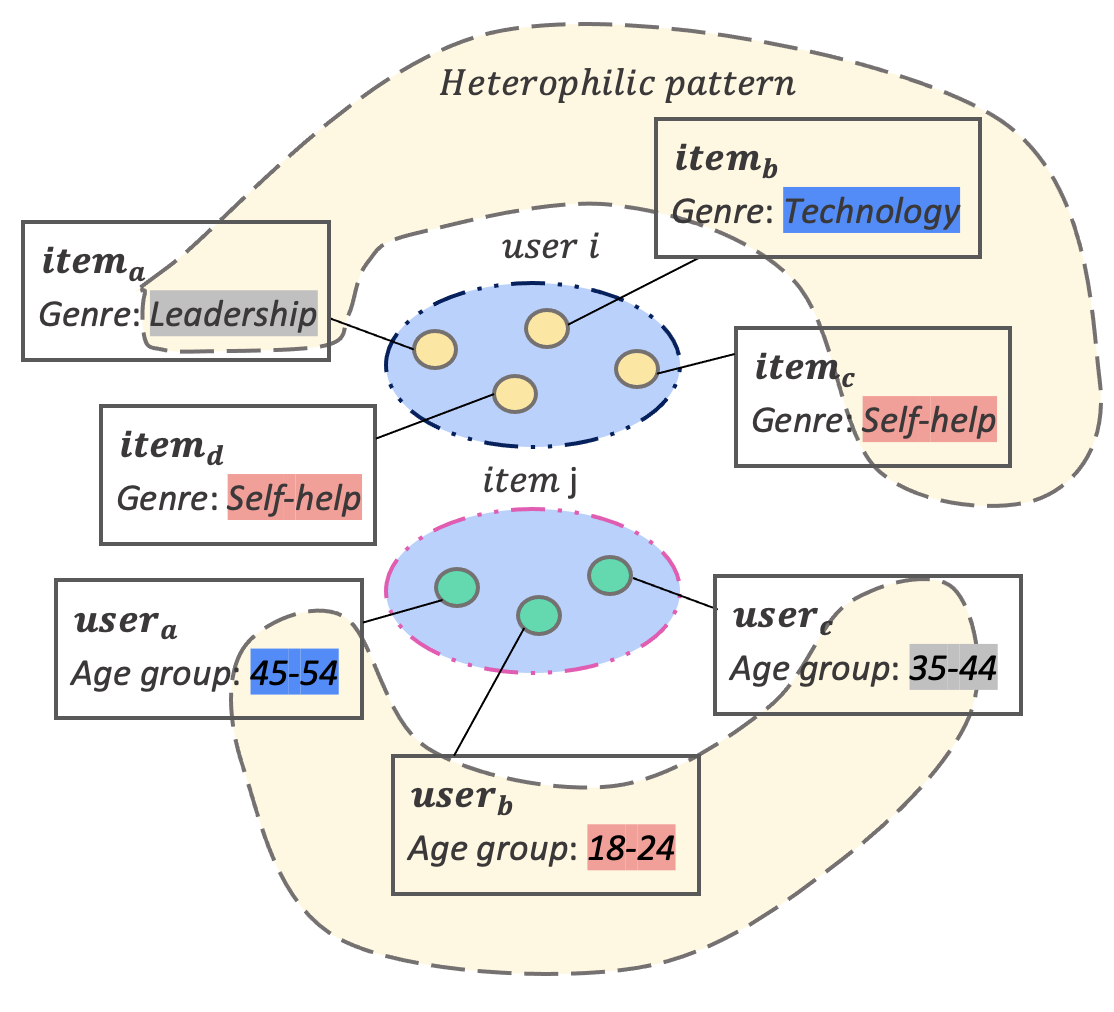}
	\caption{Example of heterophilic pattern in user and item hypergraphs.}
	\label{fig:heterophilic_example}
\end{figure}

\edit{While hypergraph-based methods have achieved notable advancements, they often neglect the presence of heterophilic patterns in user-item interaction data. 
\autoref{fig:heterophilic_example} illustrates a scenario where users and items exhibit diverse characteristics -- often referred to as a \textit{heterophilic pattern}. On the top oval, a single user $u_i$ interacts with multiple items(e.g., $item_a$, $item_b$, $item_c$, $item_d$), each belonging to different genres (e.g., Leadership, Technology, Self-help). Meanwhile, on the bottom oval, a single item $item_j$ is consumed by various users $user_a$, $user_b$, $user_c$), each from different age groups (e.g., 45-54, 18-24, 35-44).}
\edit{By highlighting these cross-cutting relationships -- one user engaging with diverse item genres, and one item appealing to diverse user demographics -- the example emphasizes the multifaceted and heterogeneous nature of user-item interactions. Such patterns make it challenging to capture all relevant and irrelevant signals (e.g., shared preferences, demographic overlaps, cross-genre interests), which is a key motivation for developing models that can effectively handle complex recommendation settings.}

In practical applications, users typically interact with items from a wide range of categories. To effectively learn representations, items within similar or related categories should have embeddings that are closely aligned, while embeddings for items in different categories should be more distinct. Furthermore, accurately modeling multi-hop neighborhood relationships is essential for capturing complex interactions. However, many existing models rely on deeply stacked \edit{Hypergraph Convolutional Networks} (HGCN) layers, which often result in over-smoothing, making the embeddings less distinguishable. Overcoming this limitation requires a more sophisticated strategy that naturally integrates local and global connections without deep stacking, while accounting for the diversity and heterogeneity in user-item interaction graphs.

To address these challenges, we propose a novel framework based on the fusion of various modality through wavelet-driven hypergraph diffusion. Specifically, our approach integrates two distinct components that are based on fusion mechanism: a Cross-Difference Relation Encoder to model heterophilic interaction patterns and a Multi-level Cluster-wise Encoder to represent localized structural relationships.

The key contributions of this work are summarized as follows:
\begin{itemize}
    \item We propose a fusion of wavelet-based hypergraph diffusion framework, called \toolname, to capture multi-faceted features while capturing both heterophilic patterns and localized topological structures through the coordinated learning of two specialized encoders.
    
    \item A \textit{Cross-Difference Relation Encoder} is introduced, utilizing an equivariant operator inspired by ED-HNN~\cite{wang2022equivariant} to effectively propagate and distinguish messages to heterogeneous nodes.
    
    \item A \textit{Multi-level Cluster-wise Encoder} is presented, combining wavelet transforms~\cite{sun2021heterogeneous} with hypergraph convolutional layers to flexibly diffuse information and robustly learn structural patterns in the hypergraph.
    
    \item To integrate features from the two encoders, we implement a multi-level fusion mechanism, incorporating intermediate fusion for partial feature interactions and late fusion to consolidate fully learned representations from each channel.
    
    \item We employ Multi-Encoder Collaborative Supervision which adopts cross-view contrastive learning to ensure consistent embeddings for the same entities across different perspectives, improving the alignment of features from the two encoders.
    
    \item Extensive experiments were conducted on real-world recommendation datasets, demonstrating the superior performance of our model compared to state-of-the-art baselines.
\end{itemize}


\section{Model and Problem Formulation}
\label{sec:model}
In this section, we begin by presenting the structural data, which includes user-item interactions and user-item hypergaph, followed by a formal definition of heterophilic pattern and our task.

\sstitle{Interaction Graph}
The interaction graph represents the explicit relationships between users and items. Formally, it is a bipartite graph \( G = (U, I, E) \), where \( U \) and \( I \) are the sets of users and items, respectively, and \( E \) is the set of edges indicating interactions (e.g., clicks, purchases, or ratings) between users and items. Each edge \( (u, i) \in E \) may be associated with additional attributes, such as timestamps or interaction strengths, which can enhance the representation of the graph.

\sstitle{User-Item Hypergraph} Hypergraphs can be divided into user hypergraph and item hypergraph. The user hypergraph \( H_u = (U, \mathcal{E}_u) \) focuses on relationships among users. Here, \( U \) is the set of user nodes, and \( \mathcal{E}_u \) is the set of hyperedges. Each hyperedge \( e \in \mathcal{E}_u \) connects a group of users sharing a common interaction pattern, such as engaging with the same item, or exhibiting similar preferences.

The item hypergraph \( H_i = (I, \mathcal{E}_i) \) focuses on relationships among items. Here, \( I \) represents the set of item nodes, and \( \mathcal{E}_i \) is the set of hyperedges. Each hyperedge \( e \in \mathcal{E}_i \) connects a subset of items that are co-interacted with by the same user or group of users. This structure enables the modeling of item co-occurrences, such as items frequently bought together or shared in playlists, and captures their interdependencies in a higher-order manner.

By combining the user hypergraph and the item hypergraph into a unified framework, the user-item hypergraph \( H = (U \cup I, \mathcal{E}) \) captures the complex and interrelated higher-order interactions between users and items.

\sstitle{Heterophilic Pattern in Hypergraph}
Heterophily refers to the tendency of nodes with different attributes or types to form connections. In the context of a user-item hypergraph, this means that users or items with contrasting characteristics may still be connected through the same hyperedge. Unlike a binary interaction graph, a hypergraph often exhibits heterophilic patterns due to its ability to model higher-order relationships, where diverse users or items with contrasting characteristics are grouped together within the same hyperedge based on shared interactions or contexts. Identifying heterophilic patterns is crucial for improving recommendation accuracy, as it allows the model to distinguish diverse and contrasting relationships within the hypergraph structure. 

\begin{mydef}[Heterophilic-aware Recommendation] 
A heterophilic-aware recommendation task aims to leverage heterophilic patterns in the user and item hypergraphs to improve the quality of recommendations. Formally, given a user \( u \in U \) and a set of candidate items \( I_c \subseteq I \), the objective is to predict a ranking \( \hat{r} \) over \( I_c \) that maximizes the likelihood of matching user \( u \)'s diverse preferences, as inferred from heterophilic relationships in the hypergraph \( H \).
\end{mydef}

\section{Methodology}
\label{sec:method}

\begin{figure}[!h]
	\centering
	\includegraphics[width=1.05\linewidth]{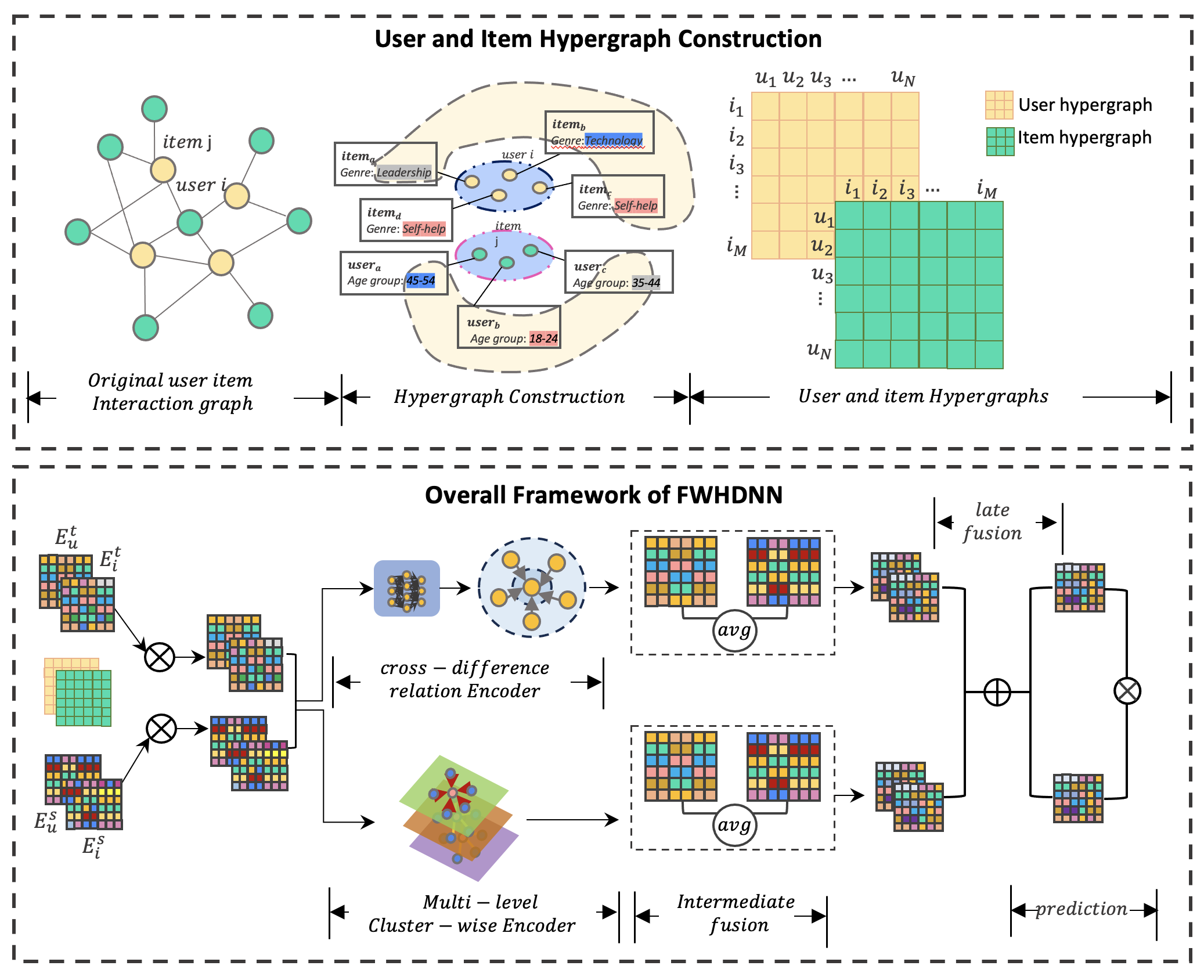}
	\caption{An illustration of \toolname framework.}
	\label{fig:framework}
\end{figure}

In this section, we first present the detailed architecture of our proposed \toolname model, as illustrated in \autoref{fig:framework}. First, we introduce the heterophily-aware hypergraph diffusion module that ensures varying message-passing for nodes in different class and labels. Second, we exploit wavelet transform-based hypergraph neural network layers to capture various scales of topological information of group-wise relationships among users and items. Third, we fuse features that capture different facets through intermediate and late fusion mechanisms to seamlessly combine structural and textual information for expressive representations. We further adopt contrastive learning to enforce the representations of equivalent entity to be close in proximity than the others in the vector space. Finally, the final user and item embeddings are harnessed to predict user interest scores on the candidate items.

\subsection{Cross-Difference Relation Encoding}
Inspired by the principles outlined in ED-HNN~\cite{wang2022equivariant}, our methodology first constructing user and item representations through the exploration of heterophilic relationships in user-item hypergraphs. In such hypergraphs, it is common to observe users grouping items into distinct categories, reflecting specific preference behaviors. To distinguish items across categories while retaining the shared characteristics within a category, we introduce a novel mechanism leveraging \textit{Heterophily-aware Hypergraph Diffusion Network (HDNN) layers}.

Our approach first applies a multi-layer perceptron (MLP) to transform the embeddings of users and items. This transformation allows the model to dynamically adapt to intricate data patterns, providing flexibility, and predefined parameter structures that may falter in scenarios where heterophilic patterns are widely spread. Afterward, the transformed embeddings are passed through HGCN, which aggregates node information within hyperedges based on topological relationships. To enhance the diversity of messages exchanged among nodes in a hyperedge and ensure equivariance, we integrate the aggregated node features with their original embeddings. This combined structure ensures that the model captures both local and global heterophilic patterns effectively. The HDNN framework can be formulated as:
\begin{align}
\mathbf{X_{e}^{(l)}} = \operatorname{LN}(\operatorname{HConv}(\operatorname{MLP}_1(\mathbf{X^{(l)}}), \mathbf{H})) + \mathbf{X^{(l)}}, \\
\mathbf{X_{v}^{(l)}} = \operatorname{LN}(\operatorname{HConv}(\operatorname{MLP}_2(\mathbf{X_{e}^{(l)}}), \mathbf{H})) + \mathbf{X_{e}^{(l)}}
\end{align}
Here, \( X \) represents the embeddings of users and items, \( X_v \) denotes the node embeddings, and \( X_e \) refers to the hyperedge embeddings. The incidence matrix is represented by \( H \), while \( \text{MLP}_1(\cdot) \) and \( \text{MLP}_2(\cdot) \) indicate the multi-layer perceptrons. The hypergraph convolution layer is denoted as \( \text{HConv}(\cdot) \), and \( \text{LN}(\cdot) \) corresponds to Layer Normalization~\cite{xu2019understanding}.
The resulting node embeddings are subsequently combined with their initial embeddings to preserve essential features captured at different layers. 

\subsection{Multi-level Cluster-wise Encoding}

Building upon the localized hypergraph neural network framework introduced in \cite{sun2021heterogeneous}, we present a wavelet-based multi-scale group-wise relationship-aware encoder designed to capture neighborhood structures across various scales while maintaining the heterogeneity inherent in hypergraphs.

The use of wavelet bases facilitates localized convolution in the vertex domain, allowing the convolution operation to focus on specific scope of regions of the hypergraph rather than treating the entire structure as a whole. This localization is especially advantageous in hypergraphs with diverse hyperedge types, as it enables the model to effectively capture the heterogeneity of the data while embedding topological information. The formulation of the wavelet-based hypergraph convolution layer is as follows:
\begin{equation}
    \mathbf{X}^{(l+1)} = \mathbf{\Theta\Lambda\Theta^{\prime}X^{(l)}W} + \mathbf{X}^{(l)}
\end{equation}
where \( \mathbf{X} \) is the feature matrix, \( \mathbf{\Lambda} \) denotes the diagonal weight filter matrix, \( \mathbf{\Theta} \) and \( \mathbf{\Theta^{\prime}} \) correspond to the wavelet transform and its inverse respectively, and \( \mathbf{W} \) represents the weight matrix for feature transformation. Note that the current node embeddings are concatenated with their newly learned representations at each layer to mitigate signal degradation across multiple layers.

\subsection{Integrated Multi-modal Fusion Mechanism}

To enhance the expressiveness and robustness of embeddings, we propose an Integrated Multi-modal Fusion Mechanism that synthesizes multi-faceted features (e.g., structural and textual features) through intermediate and late-fusion strategies. This approach leverages complementary information across modalities, enabling the construction of holistic representations that capture both shared and unique characteristics inherent to each data source. We begin by initializing the structural embeddings \edit{$\mathbf{E}_s$} to capture the intrinsic relationships among entities based on their underlying graph structure. Simultaneously, we leverage the textual embeddings \edit{$\mathbf{E}_t$} derived from user and item profiles\cite{ren2024representation}, which encode semantic information present in their descriptive attributes. 

In the intermediate-fusion stage, the two-faceted attributes are subsequently processed through the two aforementioned modules. Within each module, structural and textual embeddings are pass-forwarded independently, ensuring that two different modals are handled in a manner that preserves their unique characteristics. The features are refined and transformed to enhance their representational quality, capturing the salient aspects of their respective modalities. 
The separated outputs are then averaged, yielding a unified representation that balances the contributions of both structural and textual features.
Finally, decisions of modality-specific outputs from two modules are aggregated at a higher level in the late-fusion stage to preserve modality-specific nuances.

\subsection{Multi-Encoder Collaborative Supervision}
To ensure a unified representation of user preferences, it is essential to align the embeddings of the same users and items produced by the two separate modules, which capture distinct aspects of user behavior. This alignment helps the model integrate multiple perspectives into a consistent understanding. On the other hand, significant differences between the embeddings from the two modules might indicate redundancy or conflicting information in the captured features. To preserve the unique contributions of each encoder, the embeddings should be close to one another within the vector space.

To achieve this goal, we implement a cross-view contrastive learning approach. This method treats embeddings of the same users (or items) from different encoders as positive pairs and unrelated embeddings as negative pairs. The contrastive loss, inspired by InfoNCE~\cite{InfoNCE}, is formulated:
\begin{equation}
    \label{eqn:ssl_u}
    \mathcal{L}_s^{(u)}=\sum_{i=0}^I \sum_{l=0}^L-\log \frac{\exp \left(s\left(\mathbf{z}_{i, l}^{(u)}, \Gamma_{i, l}^{(u)}\right) / \tau\right)}{\sum_{i^{\prime}=0}^I \exp \left(s\left(\mathbf{z}_{i, l}^{(u)}, \Gamma_{i^{\prime}, l}^{(u)}\right) / \tau\right)}
\end{equation}
Here, $\tau$ represents a temperature parameter, $s(\cdot)$ denotes the cosine similarity function, $\mathbf{z}{i,l}$ corresponds to the collaborative latent vector of a user (or item) at layer $l$, and $\Gamma{i,l}$ signifies the latent vector of the same user (or item) from different encoders at the same layer.

Additionally, we incorporate the commonly used BPR loss~\cite{BPR}, which is based on the assumption that if a user $u$ interacts with an item $i$, the user is more likely to prefer $i$ over items that they have not interacted with. To optimize the model parameters, we minimize the BPR loss function:
\begin{equation}
    \mathcal{L}_{\text{BPR}} = \sum_{u \in \mathcal{U}} \sum_{i \in \mathcal{I}_u} \sum_{i' \notin \mathcal{I}_u} - \log \sigma(\hat{y}_{u,i} - \hat{y}_{u,i'})
\end{equation}
Here, $\mathcal{U}$ represents the set of users, $i \in \mathcal{I}_u$ refers to the items that a user has interacted with, and $i' \notin \mathcal{I}_u$ corresponds to items with which the user has no observed interactions. The term $\hat{y}$ indicates the likelihood of user $u$ engaging with item $i$, calculated as the dot product of the user's embedding and the item's embedding.

\section{Evaluation}
\label{sec:evaluations}

In this section, we assess the performance of \toolname by comparing it with several state-of-the-art baseline methods. We begin by outlining the experimental setup and proceed to provide a detailed performance analysis across multiple datasets. 

\subsection{Experimental Setting}
To ensure a fair comparison, we conducted evaluations on three real-world datasets: Amazon-Books for book recommendations, Steam for game recommendations, and Yelp for business recommendations. The detailed statistics for these datasets are summarized in \autoref{tab:datasets1}. These datasets were chosen to reflect varying interaction densities, showcasing the model's ability to perform robustly under diverse real-world conditions. For the initial textual embeddings, we leverage the semantic representation encoding approach proposed in \cite{wei2024llmrec} which generates textual embeddings from their profiles. Each dataset was divided into training, validation, and test sets using a 7:1:2 ratio. The final performance of each model was determined by averaging the results over five independent runs.

\begin{table}[!h]
\vspace{-1em}
\centering
\footnotesize
\caption{Statistics of Recommendation Datasets}
\label{tab:datasets1}
\begin{tabular}{c | c c c}
    \toprule
    Statistics & Amazon-Books & Steam & Yelp\\ 
    \hline 
    \#Users & 11,000 & 23,310 & 11,091 \\ 
    \#Items  & 9,332 & 5,237 & 11,010\\ 
    \#Interactions & 200,860 & 525,922 & 277,535\\ 
    Density & $1.9 \times 10^{-3}$ & $4.30 \times 10^{-3}$ & $2.2 \times 10^{-3}$\\ 
    \bottomrule
\end{tabular}
\vspace{-1em}
\end{table}

\sstitle{Metrics} 
\edit{To assess the performance of the models, we employ two widely-recognized ranking metrics: Recall@k and Normalized Discounted Cumulative Gain (NDCG@k). These metrics enable us to evaluate how effectively the models predict user interests and prioritize relevant items within the top recommendations.}

\begin{itemize}
    \item \emph{Recall@k(Recall at k):} \edit{Recall@k measures the proportion of relevant items that are successfully retrieved within the top-k recommended items. For each user, it is calculated as the number of relevant items present in the top-k list divided by the total number of relevant items available for that user. This metric provides insight into the model's ability to capture a comprehensive set of items that align with the user's interests, emphasizing the completeness of the recommendations.}
        \begin{equation}
            Recall@k = \frac{Number\_of\_relevant\_items\_in\_topk}{Total\_number\_of\_relavant\_items},
        \end{equation}
    
    \item \emph{NDCG@k:} \edit{NDCG@k evaluates the quality of the ranking of the recommended items by considering both their relevance and their positions within the top-k list. This metric assigns higher scores to relevant items that appear earlier in the ranking, reflecting the assumption that users are more likely to engage with items presented at the top of the list.
    The Discounted Cumulative Gain (DCG@k) is first calculated by summing the relevance scores of the recommended items, discounted logarithmically based on their position:}
    \begin{equation}
        \text{DCG@k} = \sum_{i=1}^{k} \frac{\text{rel}_i}{\log_2(i + 1)}
    \end{equation}
\edit{where $\text{rel}_i$ is the relevance score of the item at position $i$. NDCG@k is then obtained by normalizing DCG@k with the ideal DCG@k (IDCG@k), which is the maximum possible DCG@k for the given set of relevant items:}
    \begin{equation}
        NDCG_{@K} = \frac{DCG_{@K}}{IDCG_{@K}},
    \end{equation}
    
\end{itemize}
In the experimental settings, we specifically set the $k$ value to 10, 20, and 40.

\subsection{Baseline Models}
We evaluate the performance of our proposed model \toolname against five cutting-edge baseline models, including both GNN-based and hypergraph-based approaches for recommendation tasks.

\begin{itemize} 
    \item \textbf{LightGCN}\cite{he2020lightgcn} streamlines the NGCF framework by retaining only the neighborhood aggregation component, focusing on CF representation learning while removing unnecessary complexities. 
    \item \textbf{SGL}\cite{sgl} builds on the LightGCN framework by incorporating an augmentation strategy alongside self-supervised contrastive learning to enhance representation quality. 
    \item \textbf{DHCF}\cite{ji2020dual} employs a dual-channel hypergraph neural network to facilitate a divide-and-conquer approach, allowing separate yet simultaneous learning of user and item embeddings. 
    \item \textbf{HCCF}\cite{xia2022hypergraph} leverages contrastive learning to encode both local and global perspectives of user-item hypergraphs, enabling the model to learn distinct user and item representations effectively. 
    \item \textbf{SHT}\cite{sht2022} utilizes a transformer-based architecture paired with a hypergraph attention mechanism. Additionally, it introduces a data augmentation technique that captures diverse CF signals from various perspectives. 
    \item \textbf{AutoCF}\cite{xia2023automated} incorporates generative self-supervised learning to perform automated data augmentation, extracting meaningful self-supervised features without manual intervention. 
    \item \textbf{WaveHDNN}\cite{sakong2025high} incorporates two separate perspectives of structural information to account for heterophilic patterns and multi-scale relationships.
\end{itemize}

\subsection{Performance Comparison}
This section presents a detailed analysis of the performance results for all baselines across the three datasets.
\begin{table*}[!h]
\centering
\footnotesize
\caption{The comprehensive performance comparison of \toolname and baseline models on the Amazon-books dataset is presented. The best-performing results are emphasized in bold, while the second-best results are underlined.}
\label{tab:baseline_evaluation1}
\begin{adjustbox}{max width=0.6\linewidth}
\begin{tabular}
{@{}ccccccc@{}ccccccc@{}ccc@{}}\midrule 
\multirow{2}{*}{Model}  & \multicolumn{6}{c}{Amazon-books}\\
\cmidrule{2-7}
&Recall@10&NDCG@10&Recall@20&NDCG@20&Recall@40&NDCG@40\\ 
\midrule
\text{LightGCN} & 0.05276 & 0.04472& 0.08412 & 0.05566 & 0.13369 & 0.071 \\
\text{SGL} &0.06331 & 0.05365 & 0.10035 & 0.06664 & 0.15863 & 0.08471 \\
\text{DHCF} & 0.06355 & 0.05345 & 0.09913 & 0.06576 & 0.15285 & 0.08254 \\
\text{HCCF} & 0.07332 & 0.06133 & 0.11695 & 0.07653 & 0.17942 & 0.09599\\
\text{SHT} & 0.09123 & 0.07616 & 0.13929 & 0.09304 & 0.20774 & 0.11418 \\
\text{AutoCF} & 0.09765 & 0.0839 & 0.14391 & 0.09999 & 0.20815 & 0.12013\\
{WaveHDNN} &\underline{0.10132}& \underline{0.08657} & \underline{0.15136} & \underline{0.10389} & \underline{0.21812} & \underline{0.12484} \\
{\toolname} &\textbf{0.10368}& \textbf{0.08871} & \textbf{0.15457} & \textbf{0.10670} & \textbf{0.22382} & \textbf{0.12834} \\
\midrule
\textbf{\%Improve.} & 2.33\%& 2.47\% & 2.13\%& 2.7\%& 2.61\% & 2.81\% \\
\bottomrule
\end{tabular}
\end{adjustbox}
\end{table*}
\begin{table*}[!h]
\centering
\footnotesize
\caption{The comprehensive performance comparison of \toolname and baseline models on the Steam dataset is presented. The best-performing results are emphasized in bold, while the second-best results are underlined.}
\label{tab:baseline_evaluation2}
\begin{adjustbox}{max width=0.6\linewidth}
\begin{tabular}
{@{}ccccccc@{}ccccccc@{}ccc@{}}\midrule 
\multirow{2}{*}{Model}  & \multicolumn{6}{c}{Steam}\\
\cmidrule{2-7}
&Recall@10&NDCG@10&Recall@20&NDCG@20&Recall@40&NDCG@40\\ 
\midrule
\text{LightGCN} & 0.06503 & 0.06021 & 0.10555 & 0.07535 & 0.16418 & 0.09554\\
\text{SGL} & 0.06541 & 0.06067 & 0.10524 & 0.07571 & 0.16385 & 0.09549\\
\text{DHCF} & 0.07393 & 0.06773 & 0.11983 & 0.08515 & 0.18867 & 0.10842\\
\text{HCCF} & 0.07756 & 0.0721 & 0.12503 & \underline{0.08953} & 0.19201 & 0.11195\\
\text{SHT} & 0.08006 & 0.07583 & 0.12687 & 0.02647 & 0.19513 & 0.11579\\
\text{AutoCF} & 0.06562 & \underline{0.07632} & 0.10627 & 0.07615 & 0.16475 & 0.09567\\
{WaveHDNN} & \underline{0.08607}&\underline{0.07754}&\underline{0.13525}&\underline{0.09602} & \underline{0.20651} & \underline{0.12019}\\
{\toolname} & \textbf{0.08982}&\textbf{0.08391}&\textbf{0.15406}&\textbf{0.09900} & \textbf{0.24518} & \textbf{0.12374}\\
\midrule
\textbf{\%Improve.} & 4.36\%& 8.21\%& 13.9\% & 3.08\% & 18.72\%& 2.95\%\\
\bottomrule
\end{tabular}
\end{adjustbox}
\end{table*}

\begin{table*}[!h]
\centering
\footnotesize
\caption{The comprehensive performance comparison of \toolname and baseline models on the Yelp dataset is presented. The best-performing results are emphasized in bold, while the second-best results are underlined.}
\label{tab:baseline_evaluation3}
\begin{adjustbox}{max width=0.6\linewidth}
\begin{tabular}
{@{}ccccccc@{}ccccccc@{}ccc@{}}\midrule 
\multirow{2}{*}{Model}  & \multicolumn{6}{c}{Yelp} \\
\cmidrule{2-7}
&Recall@10&NDCG@10&Recall@20&NDCG@20&Recall@40&NDCG@40 \\ \midrule
\text{LightGCN} & 0.05379 & 0.05072& 0.08854 & 0.06302 & 0.14062 & 0.08065\\
\text{SGL} & 0.0543 & 0.05103 & 0.08839 & 0.06308 & 0.14138 & 0.08093\\
\text{DHCF} & 0.05302 & 0.04991 & 0.08855 & 0.06255 & 0.144 & 0.08154\\
\text{HCCF} & 0.06265 & 0.05825 & 0.10426 & 0.07323 & 0.16882 & 0.09505\\
\text{SHT} & {0.06955} & {0.06474} & 0.11281 & {0.08021} & 0.1836 & {0.10403}\\
\text{AutoCF} & 0.06694 & 0.06158 & {0.11401} & 0.07856 & {0.18773} & 0.10346\\
{WaveHDNN} &\underline{0.07393}& \underline{0.06831} & \underline{0.1212} & \underline{0.08502} & \underline{0.19427} & \underline{0.10971}\\
{\toolname} & \textbf{0.07644}&\textbf{0.07238}&\textbf{0.13158}&\textbf{0.08750} & \textbf{0.21216} & \textbf{0.11286}\\
\midrule
\textbf{\%Improve.} & 3.39\%& 5.95\% & 8.56\%& 2.92\%& 9.21\% & 2.88\%\\
\bottomrule
\end{tabular}
\end{adjustbox}
\end{table*}

\sstitle{Overall Comparison} 
As illustrated in \autoref{tab:baseline_evaluation1}, \autoref{tab:baseline_evaluation2}, and \autoref{tab:baseline_evaluation3}, the proposed \toolname model consistently outperforms all baseline methods across the Amazon-books, Steam, and Yelp datasets, underscoring its robustness in diverse recommendation scenarios. On the Steam dataset, which comprises user interaction data from the Steam online gaming platform, \toolname achieves a notable improvement of $2.95\%$ in NDCG@40 compared to the second-best model, highlighting its effectiveness in capturing the relevance of the top-40 recommendations. On the Amazon-books dataset, which is relatively sparse, \toolname demonstrates consistent gains, including a $2.81\%$ improvement in NDCG@40. Similarly, on the Yelp dataset -- another challenging sparse environment -- \toolname yields substantial improvements, such as $8.56\%$ in Recall@20 and $9.21\%$ in Recall@40, outperforming all baselines. These results are particularly significant because sparse datasets with limited user-item interactions often hinder representation learning, yet \toolname adapts well and delivers consistently strong performance across all benchmarks.

The superior performance of \toolname is driven by three key innovations: (1) its ability to learn distinctive representations through hypergraph convolution networks with equivariant operators, preserving both local and global neighborhood information; (2) the integration of multiple scales of localized features via a combination of hypergraph convolution and wavelet-based hypergraph transform layers, enabling richer and more expressive feature extraction; and (3) the adoption of intermediate and late fusion strategies to combine structural and textual embeddings, thereby capturing diverse data modalities. 

\sstitle{Superiority over Hypergraph-based baselines}
The experimental findings demonstrate that hypergraph-based approaches consistently surpass traditional graph-based models such as LightGCN and SGL across all datasets. This consistent performance gap underscores the benefits of using hypergraphs to represent more intricate user–item relationships compared to standard binary interaction graphs. By capturing high-order group interactions that conventional graph methods overlook, hypergraph-based models can generate more precise and robust recommendations.

Notably, SHT, a representative hypergraph-based approach, outperforms SGL by $34.8\%$, $21.2\%$, and $29.6\%$ with respect to NDCG@40 on the Amazon-books, Steam, and Yelp datasets, respectively. These substantial improvements highlight the strength of hypergraphs in modeling complex group interactions and higher-order dependencies, leading to more accurate and reliable recommendation outcomes.

Among hypergraph-based models, \toolname stands out by further pushing the performance boundaries. On the Amazon-books dataset, \toolname achieves a $2.81\%$ gain in NDCG@40 over the next-best baseline. On the Steam dataset, \toolname delivers a $2.95\%$ improvement in NDCG@40, while on the Yelp dataset, it achieves notable gains of $8.56\%$ and $9.21\%$ in Recall@20 and Recall@40, respectively. These consistent improvements across different domains -- including both dense (Steam) and sparse (Amazon-books, Yelp) datasets -- demonstrate \toolname's adaptability and effectiveness in capturing user preferences under varying levels of interaction sparsity.

In summary, the results confirm the superiority of hypergraph methods over traditional graph-based approaches, especially for modeling complex, high-order dependencies in recommendation tasks. Moreover, among hypergraph-based approaches, \toolname consistently achieves state-of-the-art performance, owing to its advanced integration of structural and textual information, multi-scale feature learning, and robust message-passing mechanisms.

\begin{table*}[!h]
\vspace{-1em}
\centering
 \footnotesize
\caption{Ablation test of \toolname (in bold) with its variants.}
\label{tab:ablation}
\begin{adjustbox}{max width=0.8\linewidth}
\begin{tabular}{@{}lcc@{\hskip 0.5cm}cc@{\hskip 0.5cm}cc@{\hskip 0.5cm}cc@{}}
\toprule
\multirow{2}{*}{Ablation Settings} & \multicolumn{2}{c}{Amazon-books} & \multicolumn{2}{c}{Steam} & \multicolumn{2}{c}{Yelp}\\
\cmidrule{2-3} \cmidrule{4-5} \cmidrule{6-7}
& Recall@40 & NDCG@40 & Recall@40 & NDCG@40 & Recall@40 & NDCG@40\\
\midrule
{\toolname} & \textbf{0.22382} & \textbf{0.12834} & \textbf{0.24518} & \textbf{0.12374} & \textbf{0.21216} & \textbf{0.11286} \\
\midrule
\text{w/o Cross-Difference Relation Encoder} & 0.18623 & 0.10627 & 0.19123 & 0.1104  & 0.17236 & 0.09499 & \\
\text{w/o Multi-level Cluster-wise Encoding} & 0.21035 & 0.11626 & 0.18845 & 0.10701 & 0.16856 & 0.09368 \\
\text{w/o Integrated Multi-modal Fusion Mechanism} & 0.21459 & 0.10489 & 0.21024 & 0.11014 & 0.17565 & 0.08448 \\
\bottomrule
\end{tabular}
\end{adjustbox}
\vspace{-1em}
\end{table*}

\subsection{Ablation Test}
In this section, we assess the significance of the key components of our \toolname. We create variation models by replacing or removing each component, demonstrating the performance improvements achieved by the proposed elements. The statistics of results are summarized \autoref{tab:ablation}.

\sstitle{Cross-Difference Relation Encoder}
This encoder enhances the model's ability to capture diverse user–item relationships by incorporating heterophily into the graph structure. As shown in Table 5, removing this component leads to a clear drop in performance across all datasets (e.g., Amazon-books: Recall@40 falls from 0.22382 to 0.18623). The encoder's adaptive message-passing mechanism within HGCN layers helps preserve diversity and mitigates over-smoothing, resulting in more robust embeddings and improved recommendation accuracy.

\sstitle{Multi-level Cluster-wise Encoding}
The wavelet-based HGCN layer captures multi-scale graph structures, ensuring that both global and fine-grained neighborhood patterns are preserved. Eliminating this mechanism causes a noticeable degradation, as reflected in the Steam dataset where Recall@40 decreases from 0.24518 to 0.18845. These results emphasize its critical role in maintaining structural detail across scales. By aggregating information at multiple levels, the model avoids over-smoothing and achieves more accurate collaborative filtering.

\sstitle{Integrated Multi-modal Fusion Mechanism}
This mechanism combines heterogeneous feature spaces, enabling the model to leverage complementary modalities for richer representation learning. As Table 5 shows, excluding this component produces consistent performance losses (e.g., Yelp: NDCG@40 drops from 0.11286 to 0.08448). The results highlight the importance of effective fusion in capturing inter-modal signals, ensuring that \toolname can fully exploit structural and contextual information. This synergy across modalities underpins the model's superior ability to generalize and deliver robust predictions.

\subsection{Hyper-parameter Studies}
This experiment examines how sensitive our model is to different hyperparameters. Among the commonly used hyperparameters, we specifically focus on the number of HGCN layers and the embedding dimension, as these factors have a notable influence on performance outcomes.

\begin{figure}[!h]
  \centering
    \includegraphics[width=1.0\linewidth]{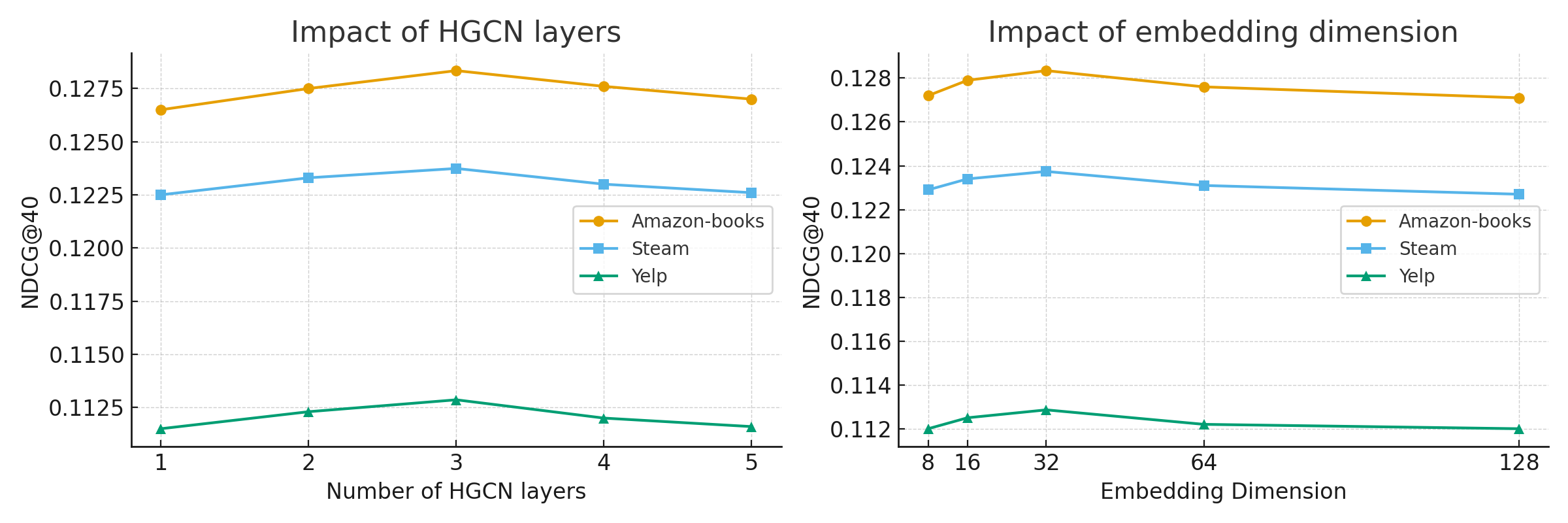}
	\vspace{-1em}
    \caption{Impact of the number of HGCN layers and different embedding size.}
    \label{fig:sensitivity}
	\vspace{-1em}
  \end{figure}

\sstitle{Impact of HGCN Layer Depth}
We evaluated the effect of varying the number of HGCN layers on \toolname's performance, as shown in \autoref{fig:sensitivity} (left). The number of layers was varied from 1 to 5 across the Amazon-books, Steam, and Yelp datasets. The results reveal a consistent trend: performance improves as the number of layers increases up to three, after which it slightly declines. For instance, Amazon-books achieves its highest NDCG@40 at three layers, while both Steam and Yelp show similar peak behavior before a mild reduction at deeper depths.

This pattern suggests that moderate depth allows the model to capture meaningful higher-order relationships without oversmoothing. However, adding more layers introduces noise from overly distant neighbors, which can dilute useful signals. These findings highlight the importance of choosing an appropriate number of layers -- in this case, three -- as a balance between local connectivity and global structural information.

\sstitle{Impact of Embedding Dimension}
We further examined the influence of embedding size, varying the dimensionality across [8, 16, 32, 64, 128]. As depicted in \autoref{fig:sensitivity} (right), performance improves as the dimension increases to 32, where the highest NDCG@40 values are observed across all three datasets. Beyond this point, accuracy remains stable or slightly decreases. For example, Amazon-books peaks at 32 dimensions, while Steam and Yelp exhibit similar trends with only marginal gains at larger sizes.

The diminishing returns at higher dimensions suggest that excessively large embeddings may spread information too thinly across dimensions, reducing the model's ability to focus on meaningful patterns. Moreover, larger embeddings impose additional computational and memory overheads without commensurate improvements in accuracy. Thus, selecting a moderate embedding size, such as 32, is optimal to ensure both efficiency and robust performance in recommendation tasks.

\section{Related Works}
\label{sec:related}

\sstitle{Traditional Collaborative Filtering}
CF is a fundamental technique in recommender systems, aiming to predict user preferences based on historical user-item interactions. CF methods can be broadly categorized into memory-based and model-based approaches.
Memory-based techniques, such as User-Based CF and Item-Based CF, leverage similarity measures like cosine similarity or Pearson correlation to estimate the likeness between users or items~\cite{breese2013empirical, sarwar2001item}. While these methods are straightforward and intuitive, they often struggle with scalability and data sparsity, particularly in large and complex real-world datasets \cite{shi2014collaborative,thang2022nature,duong2022efficient,nguyen2020factcatch,hung2017answer,nguyen2017argument}.

Model-based approaches, on the other hand, address these challenges by learning latent representations of users and items. For instance, Matrix Factorization (MF) projects users and items into a shared latent space for scalable predictions. BiasMF~\cite{koren2009matrix} extends MF by incorporating user and item bias terms, improving its ability to capture individual characteristics. Singular Value Decomposition (SVD)~\cite{koren2009matrix} decomposes the user-item interaction matrix into latent factors, enabling predictions of future interactions. An advanced variant, SVD++~\cite{koren2008factorization}, integrates implicit feedback, such as clicks or views, to capture finer user preferences. Non-Negative Matrix Factorization (NMF)~\cite{zhang2006learning} enhances MF by enforcing non-negativity constraints, ensuring that the latent factors are interpretable and positive.

To account for non-linear relationships between users and items, Neural Collaborative Filtering (NCF)~\cite{he2017neural} replaces the traditional inner product in MF with a neural network, allowing the model to learn complex user-item interactions through multi-layer feature transformations.
While MF and NCF are effective, they assume static user preferences, which may not align with real-world scenarios where preferences evolve. To address this, sequence-aware models, such as Recurrent Neural Networks (RNNs)~\cite{hidasi2015session} and Transformer-based models~\cite{kang2018self}, have been developed to model the sequential patterns of user interactions, capturing dynamic preferences over time.
Despite the emergence of more sophisticated models, foundational techniques like BiasMF and NCF remain essential in the CF community. These models strike a balance between performance and interpretability and often serve as a basis for building advanced hybrid models.

Traditional CF methods are beyond the scope of this paper's experimental comparisons. Instead, we concentrate on graph-based CF models due to their superior ability to capture complex, higher-order user-item interactions and leverage relational information effectively~\cite{he2020lightgcn,xia2022hypergraph}. This focus allows for a more comprehensive analysis aligned with our proposed hypergraph-based approach~\cite{zhao2021eires,huynh2021network,duong2022deep,nguyen2022model,nguyen2022detecting,trung2022learning,tam2023wsdm}.

\sstitle{Graph-based Collaborative Filtering}
To effectively model user preference behaviors, graph structures have become a popular approach, where users and the items they interact with are represented as nodes connected by edges~\cite{huynh2021network, duong2022deep, nguyen2023example, trung2022learning}. Traditional methods, such as ItemRank~\cite{gori2007itemrank}, assign weights to items based on relationships observed through random walks, while SimRank~\cite{jeh2002simrank} calculates user and item similarity by analyzing shared neighbors in the graph structure.

In recent years, GNNs have emerged as a powerful tool for CF, leveraging graph-based representations to capture user-item interaction patterns. For instance, NGCF~\cite{wang2019neural} enhances the propagation process within user-item graphs by extending GCNs. AGCN~\cite{feng2019attention} incorporates attention mechanisms to assign varying weights, emphasizing the relevance of nodes during propagation. LightGCN~\cite{he2020lightgcn} streamlines the GCN framework by removing non-linear activation functions and feature transformations, focusing solely on embedding propagation. UltraGCN~\cite{mao2021ultragcn} introduces an even more lightweight approach for CF, simplifying the learning of user-item representations.

PinSage~\cite{ying2018graph} scales GNNs to industrial-level recommendation systems by combining random walks with graph convolutions for efficient information propagation. Other methods have addressed common challenges in GNNs, such as over-smoothing and distinguishing relevant from irrelevant neighbors. GraphRec~\cite{fan2019graph} incorporates an attention mechanism to prioritize important neighbors during message passing, improving embedding quality. Star-GCN~\cite{zhang2019star} adopts a graph auto-encoder to generate representations for previously unseen users and items. Similarly, GCCF~\cite{chen2020revisiting} tackles over-smoothing by removing non-linearities and employing a residual network design.

Despite the progress made by graph-based CF models, existing approaches often overlook heterogeneity patterns and higher-order connections in user-item interaction graphs. To address these limitations, our proposed model introduces a novel message-passing algorithm capable of learning representations while effectively capturing high-order relationships in the graph structure~\cite{nguyen2023poisoning,nguyen2023example,nguyen2014reconciling,nguyen2015smart,thang2015evaluation,nguyen2015tag,hung2019handling}.

\sstitle{Hypergraph-based Collaborative Filtering}
Beyond binary interactions between users and items, group-level relationships frequently arise and play a central role in conveying higher-order relational information. To capture these group-wise connections, researchers have introduced hypergraph structures where any number of nodes can be connected through a hyperedge, which have proven effective at offering rich structural insights. Alongside the emergence of hypergraphs, Hypergraph Convolutional Networks (HGCNs)~\cite{feng2019hypergraph} have emerged as a powerful extension of GNNs in CF, overcoming certain limitations of traditional graph-based methods by modeling more intricate user-item interactions.

Building on the HGCN framework, DHCF~\cite{ji2020dual} adopts a divide-and-conquer strategy to simultaneously learn user and item embeddings while preserving their individual characteristics. HCCF~\cite{xia2022hypergraph} incorporates self-supervised learning into hypergraph structures, using contrastive learning to align global and local representations. This approach utilizes both node-level and hypergraph-level perspectives, enhancing the model generalization ability. Another self-supervised approach, SHT~\cite{sht2022}, introduces a hypergraph transformer network to integrate neighborhood information across multiple scales. Additionally, HypAR~\cite{jendal2024hypergraphs} delivers review-level and graph-based explainable recommendations by employing a flexible, preference-agnostic framework that leverages HGCN layers.

Meanwhile, UPRTH~\cite{yang2024unified} designs task-specific hypergraphs and employs a transitional attention layer to evaluate the relevance between a task and corresponding recommendations. MHCN~\cite{yu2021self} leverages a multi-channel HGCN to capture higher-order user relations for social recommendation, with each channel focusing on distinct relational patterns. By incorporating self-supervised learning, MHCN also restores lost connections through hierarchical mutual information maximization.

Although hypergraph-based collaborative filtering methods have achieved considerable success, many room for improvement still exists. With this motivation, our proposed model focuses on propagating permutation equivariant messages through HGCN layers, thus making embeddings more distinguishable. Furthermore, by introducing wavelet-transform-inspired mechanisms, we leverage wavelets at multiple scales to comprehensively capture diverse structural features. Through these advancements, we aim to enhance both the representational power and generalization capability of hypergraph-based CF models~\cite{yang2024pdc,sakong2024higher,huynh2024fast,huynh2025certified,nguyen2023isomorphic,nguyen2024portable}.

\section{Conclusion}
\label{sec:conclusion}

We introduce \toolname, a novel CF framework that simultaneously captures heterophilic patterns and multi-scale structural information via a wavelet transform-based hypergraph diffusion algorithm. Heterophilic user-item interaction patterns are addressed through a cross-difference relation encoder, which applies distinct message passing strategies based on node types. Additionally, multi-scale group-wise structures are incorporated via a multi-level cluster-wise encoder, leveraging wavelet-aware hypergraph convolutional networks (HGCNs). Each encoder combines textual and structural perspectives through intermediate fusion, and the final embedding is derived using a late-fusion mechanism.
Empirical results on real-world datasets demonstrate the superiority of \toolname over state-of-the-art baseline models. Future directions include extending \toolname to handle dynamic user preferences and more complex interaction types for broader applicability in recommendation systems~\cite{nguyen2024manipulating,nguyen2025privacy,pham2024dual,nguyen2024multi,nguyen2024handling}.



\end{document}